\documentclass[prl,twocolumn,showpacs,superscriptaddress]{revtex4-1}

\usepackage{amsmath}    % need for subequations
\usepackage{graphicx}   % need for figures
\begin{document}

\title{Tracking Temperature Dependent Relaxation Times of Individual Ferritin 
Nanomagnets with a Wide-band Quantum Spectrometer}
\author{Eike Sch\"afer-Nolte}
\affiliation{Max-Planck Institute for Solid State Research, 70569 Stuttgart, 
Germany}
\affiliation{3rd Institute of Physics and Research Center SCoPE, University 
Stuttgart, 70569 Stuttgart, Germany}
\author{Lukas Schlipf}
\affiliation{Max-Planck Institute for Solid State Research, 70569 Stuttgart, 
Germany}
\affiliation{3rd Institute of Physics and Research Center SCoPE, University 
Stuttgart, 70569 Stuttgart, Germany}
\author{Markus Ternes}
\email{m.ternes@fkf.mpg.de}
\affiliation{Max-Planck Institute for Solid State Research, 70569 Stuttgart, 
Germany}
\author{Friedemann Reinhard}
\affiliation{3rd Institute of Physics and Research Center SCoPE, University 
Stuttgart, 70569 Stuttgart, Germany}
\author{Klaus Kern}
\affiliation{Max-Planck Institute for Solid State Research, 70569 Stuttgart, 
Germany}
\affiliation{Institut de Physique de la Mati\`{e}re Condens\'{e}e, \'{E}cole 
Polytechnique F\'{e}d\'{e}rale de Lausanne, 
1015 Lausanne, Switzerland}
\author{J\"org Wrachtrup}
\affiliation{3rd Institute of Physics and Research Center SCoPE, University 
Stuttgart, 70569 Stuttgart, Germany}
\affiliation{Max-Planck Institute for Solid State Research, 70569 Stuttgart, 
Germany}
\date{\today}

\begin{abstract}
We demonstrate the tracking of the spin dynamics of ensemble and individual 
magnetic ferritin proteins from cryogenic up to room temperature using the 
nitrogen-vacancy color center in diamond as magnetic sensor. We employ different 
detection protocols to probe the influence of the ferritin nanomagnets on the 
longitudinal and transverse relaxation of the nitrogen-vacancy center, which 
enables magnetic sensing over a wide frequency range from Hz to GHz. The 
temperature dependence of the observed spectral features can be well understood 
by the thermally induced magnetization reversals of the ferritin and enables the 
determination of the anisotropy barrier of single ferritin molecules.
\end{abstract}


\maketitle

The study of magnetic fluctuations -- time-dependent deviations of a magnetic 
system from equilibrium -- is of high intrinsic interest and provides a powerful 
tool to gain insight into magnetic coupling mechanisms \cite{Lee06, Coffey12}. 
%For example, magnetic 
%fluctuations in high-$T_C$ superconductors have provided important information 
%to the pairing mechanism at work in their superconducting equilibrium state 
%\cite{Lee06}. 
Experimentally, however, access to fluctuations is frequently 
challenging, owing to two reasons: Firstly, fluctuations can extend over an 
excessively large range of frequencies. Secondly, many relevant magnetic systems 
are small entities with dimensions in the nanometer range, such as single 
molecules, clusters, or magnetic domains. Hence, the study of magnetic 
fluctuations requires a measurement technique featuring simultaneously nanoscale 
resolution and a wide frequency bandwidth.

Widely used methods to study ensembles of magnetic nanosystems are SQUID 
magnetometry \cite{Gider95, Gilles02}, M\"o\ss bauer spectroscopy 
\cite{Papaefthymiou10}, electron spin resonance \cite{Wajnberg01}, neutron 
scattering \cite{Morup07}, or X-ray magnetic circular dichroism 
\cite{Gambardella03}. Furthermore, individual magnetic nano-objects have been 
investigated by scanning probe microscopy \cite{Loth12, Khajetoorians13}, 
micro-SQUIDs \cite{Wernsdorfer07, Vohralik09} or scanning X-ray microscopy 
\cite{Balan14}. In general these techniques have a limited detection 
bandwidth, such that the experimental observation of magnetic fluctuations in a 
wide frequency range relies on stitched measurements combining complementary 
techniques in different frequency domains. 

Here we show that the Nitrogen-Vacancy (NV) center in diamond employed as a 
magnetic field sensor \cite{Gruber97, Taylor08} simultaneously provides 
access to both nanoscale objects and a frequency bandwidth spanning ten orders 
of magnitude. We use the unique properties of the NV to monitor thermal 
magnetization reversals of single biological nano-magnets from cryogenic 
%(4~K) 
to room temperature 
%(300~K) 
where these fluctuations accelerate from the sub-Hz to the GHz range. 

\begin{figure}[]
\includegraphics[%trim = 1.9cm 20cm 8.5cm 1.3cm, clip, 
width=0.9\columnwidth]{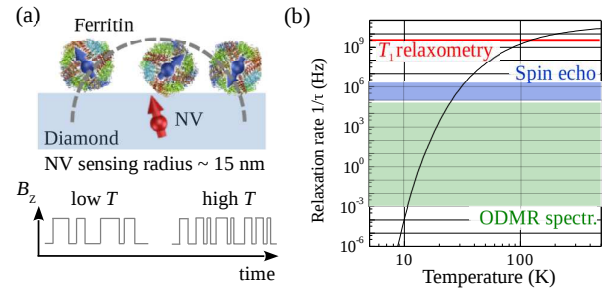}
\caption{(a) Ferritin molecules adsorbed to the diamond surface couple to 
shallow implanted NV centers. The magnetization reversals of the ferritin 
generate magnetic telegraph noise with a temperature dependent fluctuation rate. 
(b) Ferritin relaxation rate as a function of temperature for $E_a=29$~meV. 
The sensitivity ranges of different detection protocols for NV magnetometry are 
indicated.}
\label{fig1}
\end{figure}

We study ferritin protein complexes adsorbed onto a diamond surface 
(Fig.~\ref{fig1}a). Each of these proteins encloses a cluster of up to 4500 Fe 
atoms wich net magnetic moment exhibits strong thermally activated 
fluctuations. We detect the magnetic stray field of these clusters with NV 
defect centers embedded 5--10~nm below the diamond surface. This center enables 
the precise determination of the local magnetic field via the Zeeman shift of 
its spin sublevels, which can be measured by optically detected magnetic 
resonance (ODMR) techniques. Due to its atomic size, an individual NV center can 
be placed as sensor spin in nanometer proximity to the sample \cite{Taylor08, 
Degen08a}, allowing for coupling to only a single ferritin complex. Moreover, 
it can sense magnetic field fluctuations on various frequency scales from the 
sub-Hz \cite{Dreau11} over the kHz--MHz \cite{Staudacher13, Mamin13} to 
the GHz range \cite{Steinert13}, depending on the spectroscopy protocol 
employed. As a notable extension of previous studies, we here demonstrate 
nanosensing in a variable temperature setup between 5--300 K in ultrahigh vacuum 
\cite{Schaefer14}. 

\begin{figure}[]
\includegraphics[%trim = 1cm 7cm 3.5cm 8cm, clip, 
width=0.8\columnwidth]{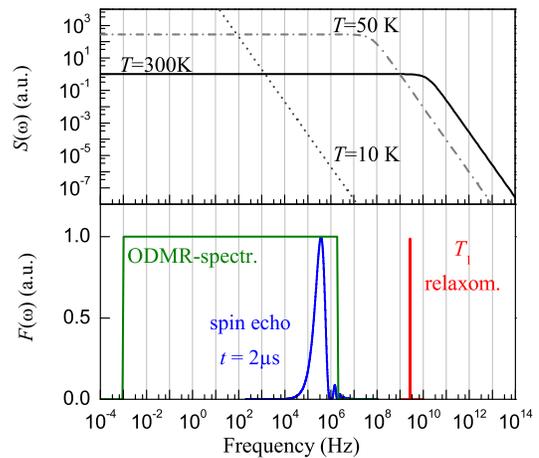}
\caption{Spectral density $S(\omega)$ of the ferritin spin noise at different 
temperatures for an anisotropy barrier of $E_a = 25$~meV. The filter functions 
$F(\omega)$ of $T_1$ relaxometry, spin echo spectroscopy and ODMR-spectroscopy 
are depicted below.}
\label{fig2}
\end{figure}

Over this temperature range, magnetic fluctuations of ferritin cores exhibit a 
rich variety of dynamic phenomena like superparamagnetism \cite{Frankel91}, 
superantiferromagnetism \cite{Gilles02}, and macroscopic quantum tunneling 
\cite{Gider95}.The Fe atoms in the biomineral core are stored in form of a 
hydrous ferric oxide-phosphate mineral similar in structure to ferrihydrate 
\cite{Bauminger89}. At lower temperatures the Fe$^{3+}$ spins are 
antiferromagnetically coupled with a Ne\'{e}l temperature estimated to be 
between 240~K \cite{Bauminger89} and 500~K \cite{Gilles02}. In the 
antiferromagnetic phase, the ferritin core possesses a net magnetic moment on 
the order of 300$\mu_B$ due to an imperfect compensation of the Fe$^{3+}$ 
spins 
on the two sublattices \cite{Harris99, Papaefthymiou10}. The core is 
generally considered as single magnetic domain with uniaxial anisotropy, 
therefore the magnetic moment fluctuates thermally activated between two easy 
directions with a temperature dependent rate. The dynamic of these stochastic 
magnetization reversals is characterized by the spin lifetime
\begin{equation}
\tau(T, E_a) = \tau_0 \cdot \exp (E_a/{k_B T})
\label{Neel Arrhenius equation}
\end{equation}
with the inverse attempt frequency $\tau_0 \approx 2 \times 10^{-11}$~s 
\cite{Morup07} and the anisotropy barrier $E_a$. This anisotropy barrier is 
dominated by the magnetocrystalline energy 
%and can be written as $E_a \approx K 
%\, V$, where $V$ is the particle Volume and $K$ is the corresponding energy 
%density 
\cite{Papaefthymiou10}. The  magnetization reversals of an individual 
molecule with the temperature dependent relaxation rate $1/\tau$ 
(Fig.~\ref{fig1}b) generates magnetic field fluctuations resembling random 
telegraph noise. The normalized spectral density of this spin noise is  
calculated to
\cite{Steinert13}
\begin{equation}
S(\omega, T, E_a)=\frac{2}{\pi} \: \frac{\tau(T, E_a)}{1+\tau(T, E_a)^2 
\omega^2}.
\label{ferritin_noise_spectrum}
\end{equation}
With lower temperatures the cut-off frequency of $S(\omega, T, E_a)$ decreases, 
while the low frequency amplitude increases (Fig.~\ref{fig2}). To account for 
the spread of $E_a$ in the measured ferritin ensembles we assume a log-normal 
distribution function \cite{Madsen08}.

Our method to detect these fluctuations is based on changes in the longitudinal 
($T_1$) or transverse ($T_2, T_2^{*}$)
\footnote{$T_2^{*}$ is the spin decoherence time affected mainly by 
low-frequency noise compared to $T_2>T_2^{*}$ which is only sensitive to noise 
in the detection window at higher frequency.} 
spin relaxation time of the NV center in response to the ferritin 
spin noise. The relaxation rate of the NV center can be written in general as 
\begin{equation}
\frac{1}{T_i}= \left(\frac{1}{T_i}\right)_{\rm int} + \int \left\langle 
B^2\right\rangle \cdot S(\omega, T, E_a) \cdot F_i(\omega) \: d\omega,
\label{relaxation rate int}
\end{equation}
where $\left( T_i^{-1} \right)_{\rm int} $ accounts for intrinsic relaxation 
mechanisms and $\sqrt{\left\langle B^2 \right\rangle}$ is the effective magnetic 
field at the NV position generated by the magnetic ferritin core, corresponding 
to the distance-dependent dipolar coupling strength. In this equation the 
spectral response function $F_i(\omega)$ depends on the employed sensing 
protocol. To access different frequency ranges, we probe the 
%$T_1$, $T_2$, and $T_2^{*}$ 
relaxation of the NV center by the inversion recovery protocol 
\cite{Jarmola12}, spin echo spectroscopy \cite{Laraoui10}, and 
ODMR-spectroscopy \cite{Dreau11}. The corresponding response functions and 
spectral sensitivity ranges are summarized in table \ref{table filter functions} 
and Fig.~\ref{fig2}. For ODMR spectroscopy, which was carried out by applying a 
single microwave pulse with varying frequency, we assumed a constant sensitivity 
in a frequency window limited by the averaging time per spectrum $t_{ avg}= 
10^3$~s and the length of the microwave pulse $t_{\pi} = 500$~ns. 

\begin{table}
\label{table1}
\begin{ruledtabular}
\begin{tabular}{lcc}
Technique & Filterfunction $F_i(\omega)$ & Frequency range\\
\hline \\
$T_1$ relaxom. & $\displaystyle \frac{1}{\pi} \: 
\frac{\Gamma}{{\Gamma}^2+(\omega-\omega_L)^2}$ & $\sim$ 3 GHz  \\ 
Spin echo & $\displaystyle \frac{1}{t} \left|\frac{\sin^2(\omega t/4)}{\omega 
/4}\right|^2$ & 0.1--1 MHz\\
ODMR spectr. & $\left\{ 
\begin{array}{cc}
    \displaystyle \frac{1}{c}, & \: \displaystyle \frac{2 \pi}{t_{acq}} < 
\omega < \displaystyle \frac{2 \pi}{t_{\pi}}\\
    0, & \: \text{else} 
\end{array} \right.$ & $10^{-3}$--$10^6$ Hz \vspace*{1ex} \\
\end{tabular}
\caption{Filter functions $F(\omega)$ and corresponding frequency ranges for 
the employed detection protocols. $\Gamma=1/T_2^{*}$: NV dephasing rate, 
$\omega_L$: NV transition frequency, $t$: free evolution time, $c$: 
normalization constant, $t_{acq}$: acquisition time, $t_{\pi}$: microwave pulse 
length.}
\label{table filter functions}
\end{ruledtabular}
\end{table}

\begin{figure*}[t]
\includegraphics[%trim = 1.5cm 22.5cm 3cm 2.5cm, clip, 
width=0.85\textwidth]{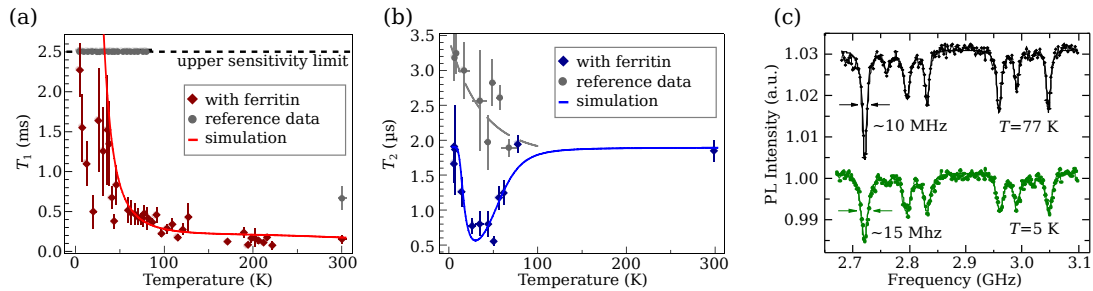}
\caption{Measurements on NV ensembles interacting with ferritin molecules. (a) 
Longitudinal relaxation time $T_1$ (diamonds), and simulation using 
$\overline{E_a} =15$~meV and $\sqrt{\left\langle B^2\right\rangle} = 22$~MHz 
(full line). The longest 
detectable decay time of 2.5~ms is indicated as upper sensitivity limit by the 
dashed line. (b) Spin coherence time $T_2$ (diamonds), and simulation using 
$\overline{E_a} =25$~meV and $\sqrt{\left\langle B^2\right\rangle} = 10$~MHz 
(full line). 
Reference data without ferritin in (a) and (b) are shown as circles. (c) 
ODMR spectra acquired at 5 
K and 77 K.}
\label{fig3}
\end{figure*}

With decreasing temperature the ferritin noise spectrum is subsequently shifted 
through these distinct sensitivity windows which we verify by measurements on an 
ensemble of ferritin molecules (Fig.~\ref{fig3}). Without adsorbed ferritin we 
measure $T_1=0.67\pm0.15$~ms at $T=300$~K which increases to $>2.5$~ms at low 
temperature, i.\,e.\ above the longest decay time detectable in our setup. Upon 
the adsorption of ferritin molecules, the $T_1$ time at room temperature is 
reduced by approximately a factor of 5, which has been previously observed for 
bulk diamonds \cite{Ziem13} and diamond nanocrystals \cite{Ermakova13}. The 
ferritin-induced reduction in $T_1$ vanishes at low temperatures, since the 
cut-off frequency of the ferritin noise spectrum shifts below the NV transition 
frequency. In the low temperature regime ($k_B T \ll E_a$) the ferritin 
magnetization is static on the timescale of the measurement, since magnetization 
reversals over the energy barrier are suppressed. The data show thus a strong 
increase in $T_1$ below the blocking temperature of $T_B \approx 50$~K. Assuming 
a temperature dependent intrinsic relaxation rate as found in reference 
\cite{Jarmola12}, we can well describe the measurement within our model with 
the fit-parameters $\overline{E_a} \approx 15$~meV and $\sqrt{\left\langle 
B^2\right\rangle} = 22$~MHz \footnote{The effective magnetic field as a measure 
of coupling strength is given in frequency units for a gyromagnetic ratio 
$\gamma = 28$ MHz/mT}. 

To probe the temperature dependence of the ferritin magnetization dynamics in 
the range 0.1--1~MHz we employ spin echo spectroscopy. The coherence time 
$T_2$ time is rather unaffected by the ferritin spin noise at 300~K, due to the 
low noise amplitude in the probed frequency range (Fig.~\ref{fig3}b). The 
intrinsic relaxation rate is comparably high due to the surface proximity of the 
NV centers \cite{Rosskopf14}. This changes at low temperatures; as the cut-off 
frequency of the ferritin noise spectrum decreases, the noise amplitude in the 
probed frequency range initially increases. The ferritin contribution to the 
relaxation rate becomes dominant for $T\lesssim 80$~K, leading to a reduction of 
$T_2$. At $T\lesssim 35$~K, the cut-off frequency of the ferritin noise spectrum 
 shifts below the detection window of the spin echo sequence, resulting in a 
recovery of the $T_2$ time. The minimum in $T_2$ at roughly 35~K corresponds to 
the blocking temperature for the probed frequency range. The reference 
measurement acquired without adsorbed ferritin shows only a modest increase of 
$T_2$ with lower temperature, which might be related to temperature dependent 
intrinsic relaxation processes. We reach an overall good agreement when fitting 
the data with the above described model, yielding $\overline{E_a} = 25$~meV and 
$\sqrt{\left\langle B^2\right\rangle} = 10$~MHz.

The deviation in $\overline{E_a}$ and $\sqrt{\left\langle 
B^2\right\rangle}$ between the $T_1$ and $T_2$ measurement can be rationalized 
by the fact that different samples were used, and correspondingly the 
distribution of anisotropy barriers and the number of detected molecules were 
different in both experiments. In general the obtained values correspond well to 
reported average anisotropy barriers in ferritin of $\sim 28$~meV determined by 
M\"o\ss bauer spectroscopy and magnetization measurements \cite{Madsen08, 
Kilcoyne95}.

We obtain further insight into the ferritin spin dynamics especially at low 
frequencies by ODMR spectroscopy at $T=77$~K and 5~K (Fig.~\ref{fig3}c). At 5~K 
we detect an increase of the resonance line width by roughly 5~MHz, which can be 
attributed to inhomogeneous broadening \cite{Wajnberg01}. 
At 77~K the majority 
of the ferritin molecules fluctuate with a rate faster than the inverse 
microwave pulse length (500 ns), such that only the average magnetization of the 
molecular ensemble is detected. At 5~K the spin dynamics of most molecules is 
blocked, therefore the magnetization is static over the measurement time ($\sim 
10^3\,$s)  and the local magnetic field differs for each NV center depending on 
the size and the orientation of the nearby molecules. The low temperature line 
width of $\sim$15 MHz is comparable to the coupling strength $\sqrt{\left\langle 
B^2\right\rangle}$ estimated from the $T_1$ and $T_2$ measurements. This 
corresponds to an effective internal field in the NV center ensemble of $\sim 
500\ \mu$T which is consistent with the expected stray field of ferritin 
molecules ($\mu \approx 300 \mu_B$) on the diamond surface.

The obtained blocking temperatures can be compared to values reported for 
horse-spleen ferritin using other detection techniques; susceptibility 
measurements (characteristic measurement time $\tau_m \sim 
10^{-3}\mbox{--}10^{2}\,$s) yield blocking temperatures in the range 5--20~K 
\cite{Papaefthymiou10, Harris99, Gider95, Kilcoyne95}, M\"o\ss bauer 
spectroscopy ($\tau_m \sim 10^{-8}\,$s) 30--50~K \cite{Papaefthymiou10, 
Frankel91, Bauminger89}, and electron spin resonance spectroscopy ($\tau_m 
\sim 10^{-10}\,$s) $\sim$ 100 K \cite{Wajnberg01}. Our findings using NV 
magnetometry are thus consistent with previously reported results. Furthermore, 
we point out that NV magnetometry can cover a significantly wider frequency 
range than the above mentioned techniques and that by using different detection 
protocols or applying magnetic fields the spectral sensitivity can be further 
extended.

An important advantage of the NV sensor is the extremely high sensitivity 
with the potential for the investigation of single molecules \cite{Taylor08, 
Degen08a}. To utilize this capability we reduced the coverage of the adsorbed 
ferritin such that isolated molecules are obtained. By employing individual NV 
centers as sensors, the number of detected molecules can be expected to be very 
low due to the short range interaction. We studied the same NV centers 
($\sim$~35 in total) before and after ferritin deposition and found for roughly 
50\% a pronounced reduction of $T_1$ at room temperature. These centers 
interacting with only one or a few molecules where investigated at variable 
temperature.

\begin{figure}[]
\includegraphics[%trim = 3.2cm 18cm 9cm 3cm, 
clip,width=0.9\columnwidth]{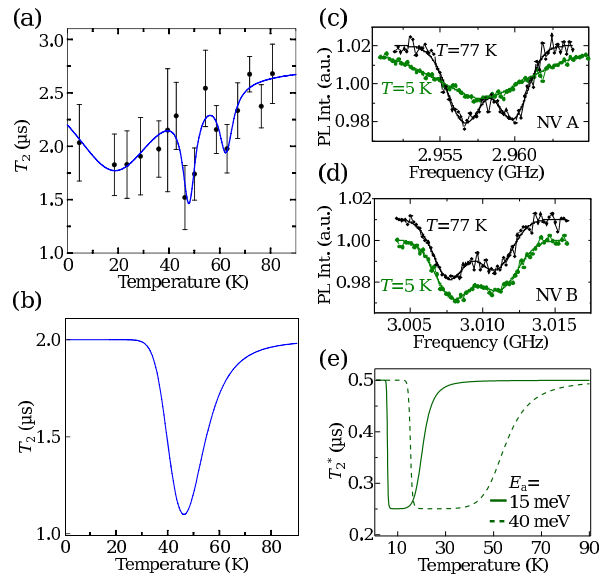}
\caption{(a) Temperature dependent $T_2$ of a single NV center interacting with 
isolated molecules. (b) Simulated $T_2$ for a NV center coupling to a single 
molecule with $E_a = 43$~meV and $\sqrt{\left\langle B^2\right\rangle} = 1.5$ 
MHz. 
(c) and (d): Representative ODMR spectra of single NV centers. Low temperature 
line broadening as shown in (a) was observed for $\sim 20$\% of the 
investigated NV centers. (e) Calculated temperature dependence of $T_2^{*}$ for 
an NV center coupling to ferritin molecules with large ($E_a= 40$~meV, full 
line) and small ($E_a =15$~meV, dashed line) anisotropy barrier.}
\label{fig4}
\end{figure}

Figure \ref{fig4}a shows the coherence time $T_2$ for a 
representative single NV center revealing several minima.
Measurements on other NV centers yield similar curves with one or multiple 
minima with a rather small width. The dips in $T_2$ fall in the same temperature 
range as the broad feature observed in the ensemble measurements 
(Fig.~\ref{fig3}b). To model the response of the NV center to a single molecule 
we use $S(\omega)$ without ensemble broadening and found a good agreement to the 
observed minimum in $T_2$ at $T= 47$~K using $E_a= 43$~meV and 
$\sqrt{\left\langle B^2\right\rangle} = 1.5$~MHz (Fig.~\ref{fig4}b). The width 
of the simulated minimum is on the order of 10~K, which corresponds well to the 
sharp features observed in the experiment. These findings support the 
interpretation that the observed minima are the fingerprint of individual 
molecules with different blocking temperatures in the vicinity of the NV center. 
This technique enables thus the determination of anisotropy barriers on the 
single molecule level.

The low temperature resonance line width in the ODMR spectra of single NV 
centers depends on the anisotropy barrier of the nearby molecules. We observe a 
broadening at 5 K for roughly 20\% of the investigated NV centers (Fig. 
\ref{fig4}c) which we attribute to the magnetic fluctuations of molecules with 
rather low anisotropy energy such that $T_B <5$~K. In contrast, most NV 
centers show similar line shapes at 5~K and 77~K (Fig.~\ref{fig4}d), indicating 
that nearby molecules are blocked at 5~K. The simulated NV decoherence time 
$T_2^{*}\propto\Gamma^{-1}$ (Fig.~\ref{fig4}e) suggests 
that the low temperature broadening of the linewidth $\Gamma$ 
is related to molecules with $E_a<15$~meV, which is consistent with the expected 
spread of anisotropy energies in the sample \cite{Madsen08}.

The static magnetization of the blocked ferritin molecules is expected to 
result in a random shift of the transition frequency between 5~K and 77~K. 
Experimentally we find this shift to be $\sim 1$~MHz for all NV centers studied, 
which is significantly smaller then the line broadening of $\sim 5$~MHz observed 
for the unblocked molecules (Fig.~\ref{fig4}c). This deviation in the coupling 
strength might be related to the existence of a second magnetic phase located at 
the particle surface \cite{Brooks98, Wajnberg01}, which aligns anti-parallel 
to the interior of the core at low temperature and thus reduces the net magnetic 
moment.

In conclusion, we demonstrated the detection of the temperature dependent 
relaxation dynamics of ferritin molecules by employing NV magnetometry. The main 
advantages of this approach are the wide frequency range that can be covered and 
the extremely high sensitivity enabling single molecule experiments. While in 
the single NV center experiments demonstrated here the number of detected 
ferritin molecules could only be estimated, future experiments using NV spin 
sensors in a scanning probe architecture \cite{Maletinsky12, Rondin12, 
Schaefer14} will 
facilitate single molecule investigations with enhanced control and precision. 
Due to its high sensitivity and 
its intrinsic spatial resolution the technique can also be applied to 
investigate systems with smaller magnetic moment, such as molecular magnets, 
radicals, or, ultimately nuclear spins.

\begin{acknowledgments}
F.R. and J.W. acknowledge financial support by the EU (Squtec, Diadems), Darpa 
(Quasar), 
BMBF (CHIST-ERA) and contract research of the Baden-W\"urttemberg foundation.
\end{acknowledgments}

%\bibliography{bib}

\end{document}